\documentclass{elsarticle}
\usepackage{lipsum}
\usepackage[inline]{enumitem}
\usepackage{enumerate}
\usepackage{multirow}
\usepackage{amsmath}
\makeatletter
\def\ps@pprintTitle{%
 \let\@oddhead\@empty
 \let\@evenhead\@empty
 \def\@oddfoot{}%
 \let\@evenfoot\@oddfoot}
\makeatother
\usepackage{hyperref}

\journal{Journal of \LaTeX\ Templates}

\begin{document}

\begin{frontmatter}

\title{Evaluating the Performance of Clone Detection Tools in Detecting Cloned Co-change Candidates}

\author{Md Nadim, Manishankar Mondal, Chanchal K. Roy, and Kevin A. Schneider}
\address{Department of Computer Science, University of Saskatchewan, Saskatoon, Canada}

\begin{abstract}
Co-change candidates are the group of code fragments that require a change if any of these fragments experience a modification in a commit operation during software evolution. The cloned co-change candidates are a subset of the co-change candidates, and the members in this subset are clones of one another. The cloned co-change candidates are usually created by reusing existing code fragments in a software system. Detecting cloned co-change candidates is essential for clone-tracking, and studies have shown that we can use clone detection tools to find cloned co-change candidates. However, although several studies evaluate clone detection tools for their accuracy in detecting cloned fragments, we found no study that evaluates clone detection tools for detecting cloned co-change candidates. In this study, we explore the dimension of code clone research for detecting cloned co-change candidates. We compare the performance of 12 different configurations of nine promising clone detection tools in identifying cloned co-change candidates from eight open-source C and Java-based subject systems of various sizes and application domains. A ranked list and analysis of the results provides valuable insights and guidelines into selecting and configuring a clone detection tool for identifying co-change candidates and leads to a new dimension of code clone research into change impact analysis.
\end{abstract}

\begin{keyword}
Clone Detection; Cloned Co-change Candidates; Commit operation; Software Maintenance and Evolution
\end{keyword}

\end{frontmatter}


\section{Introduction}
\label{introduction-cochage}
Version control systems maintain a history of changes required to keep a software system updated with the changing requirements in its life cycle \cite{VersionControl, ImpactSoftwareRequirementChange}. 
Software developers make changes to a software system by applying commit operations through its version control system, and the commits may be related to each other or maybe independent \cite{Mondal:Co-change-recommendation, Mondal:Connectivity:co-changed}. Each of such commits may perform one or more changes in the source code of the software system based on the available change requests. A set of change requests may be for different purposes, such as addressing evolving requirements or fixing bugs (e.g., problems or issues). A single commit operation may contain both related and independent changes.  Related changes are known as co-change candidates in the literature \cite{Mondal-2014-PRC-2597073-2597104rankingCoChange}, which represents a group of changes. Suppose any code fragment of a co-change group is experiencing any update. In that case, all the other code fragments in that group might need to be updated to ensure the consistent evolution of a software system \cite{Mondal:Association:Rules, Mondal:Context:Adaptation:Bugs}. When a code fragment that should co-change with its target fragment is missed or not identified, it may induce bugs or inconsistencies in a software system \cite{Judith:Bug:Replication, Judith:Micro:Regular:Clone}.

We divide the source code fragments from a group of co-change candidates into two categories. First, those fragments that are clones of one another make a group named Cloned Co-change Candidates (CCC). Second, all the other code fragments are dissimilar co-change candidates (DCC), which might be distinct enough but still need to co-change because they are functionally dependent or coupled. Therefore, both the cloned (CCC) and dissimilar (DCC) co-change candidates are the subset of the group of co-change candidates. We identify the Cloned Co-change Candidates (CCC) using different clone detection tools (listed in Table \ref{tab:tool-configurations}) by utilizing clone groups or clone classes in the detected result sets. In this study, we evaluate the performance of clone detection tools based on detecting these Cloned Co-change Candidates (CCC). 

Finding the co-change candidates (both CCC and DCC) of a target code fragment is also known as change impact analysis \cite{book-change-impact} in the literature.  \citet{Mondal:Association:Rules} investigated whether a clone detection tool can enhance the performance of an evolutionary coupling based tool in finding a change impact set or co-change candidates. Their investigation used Nicad to detect both the regular clones and micro-clones and they found that using clone information significantly enhances the performance of Tarmaq \cite{TarmaqChangeImpact}, an association rule mining based change impact analysis tool. Since they only used Nicad, we wanted to compare some other promising clone detection tools to find out whether these tools  perform better for detecting cloned co-change candidates. As well, \citet{CLEVER-JIT} used the Nicad clone detection tool to recommend qualitative fixes to developers on how to fix risky commits (commits that create inconsistencies in the system) for 12 Ubisoft systems. They first identified risky commits using a Random Forest Classifier \cite{RandomForestAlgo} based detection model and then used the Nicad clone detection tool to find similar commits with a fix already available in the history of the software system. Then they recommended the best-selected fixes to the software developer for fixing the identified risky commit. Their study showed that at least one Ubisoft software developer accepted 66.7\% of their recommended fixes. Although their study focused on one specific commercial software system and its developers, we believe clone detectors could also contribute to finding similar buggy commits and their fixes in other commercial and open-source software systems. These studies motivate us to compare 12 clone detection techniques and the findings of our study suggests important guidelines for selecting clone detectors for conducting change impact analysis. Our study's outcomes will help with the successful integration of the best performing clone detection tools with change impact analysis tools to identify risky commits and possible fixes during commit operations.   

The comparative performance of different clone detection tools in identifying cloned co-change candidates has yet to be investigated. Detecting cloned co-change candidates is useful during software maintenance to help ensure a system is changed consistently. In addition to determining whether a clone detection tool performs well in detecting clone fragments, it would be useful to know if it performs well in detecting cloned co-change candidates. 

Clone detectors combine similar code fragments that meet certain similarity thresholds (e.g., 70\% similarity) into clone groups. A clone group may contain two or more similar code fragments known as clone pairs (if the group contains exactly two similar code fragments) or clone classes (if the group contains two or more similar code fragments). Since code fragments in a clone group are similar, the clones also have similar functionality, which implies that if we want to change any of the code fragments in a clone group, the other fragments in that group may also require a similar change (co-change) to maintain consistent behavior in the software system. This assumption leads us to the possibility that all clone class members could be cloned co-change candidates of each other, and whenever any one of those fragments are modified, the developer might make a similar modification to all the other fragments of that class. We use clone classes and pairs extracted from subject systems using different clone detection tools to predict cloned co-change candidates.

We evaluated four different configurations of CloneWorks \cite{CloneWorks-Jeff} and eight additional clone detectors in our investigation. 
Each configuration of CloneWorks implements a unique mechanism for processing software system source files to detect clone fragments, and so we consider the four different configurations to be different tools.
Therefore, we evaluated 12 separate implementations of clone detection tools. 
We apply these tools to thousands of commit operations from the evolutionary histories of eight open-source software systems that vary in source code size and are from different application domains.
The clone detection tools and configurations we used in this study are shown in Table \ref{tab:tool-configurations} and the subject software systems are reported in Table \ref{tab:subject-systems-cc}.

Based on this study, we answered the following research questions:

\vspace{0.15cm}
\noindent
\textbf{RQ1: }How can we compare different clone detection tools based on the performance in detecting cloned co-change candidates? 
 
\vspace{0.15cm}
\noindent
\textbf{RQ2: }What are the deciding factors for the performance variance of different clone detectors in detecting cloned co-change candidates?

\vspace{0.15cm}
\noindent
\textbf{RQ3: }Do the source code processing techniques (Pattern/Token/Text-based processing) of the clone detection tools have any impact on their performance in detecting cloned co-change candidates?

\vspace{0.15cm}
\noindent
\textbf{RQ4: }Do clone detection tools designed for detecting different types of clones (Type 1, 2, 3) work differently in detecting cloned co-change candidates?

To the best of our knowledge, there is no other study comparing clone detection tools considering a particular maintenance perspective (such as their performance in detecting cloned co-change candidates during software evolution). One can assume that a clone detector, which is good in detecting cloned fragments \cite{Roy09comparisonand, jeff-evaluating, 4288192Comparison, ScenarioBasedComparison}, might also be good in detecting cloned co-change candidates. We wanted to verify this assumption in this exploratory study. We selected 12 implementations of clone detectors detecting different types of clone fragments to evaluate their performance in detecting cloned co-change candidates. Our investigation and analysis found that the clone detectors that detect Type-3 clones and perform pattern-based source code processing are significantly better in detecting cloned co-change candidates. Our investigation also shows that tools that provide more clone fragments in their clone detection result and cover more source code lines in the software codebase are more suitable for detecting cloned co-change candidates.

We organized this paper in the following sections: we describe related work in Section \ref{the-related-works}, our methodology is in Section \ref{the-methodology}, we describe the experimental result in Section \ref{the-experimental-result}, the discussion of the research results is in Section \ref{the-discussion}, Section \ref{the-threat-validity} explains some possible threats to validity, and we conclude our paper with some future directions of this study in Section \ref{the-conclusion-cochange}.

This paper is a significant extension of our previous work \cite{nadim-iwsc-2020} on detecting cloned co-change candidates using different clone detectors. Our previous work answered two research questions by analyzing six clone detectors on six open-source software systems. Our earlier study's two research questions showed that even though a tool that is good in detecting clone fragments from software systems may not be useful in detecting cloned co-change candidates. The tools that detect a large number of clone fragments and cover more unique lines in the source files are found suitable in predicting cloned co-change candidates. We extend our previous work by answering two additional research questions (RQ3, RQ4) to find more specific reasons for the variation of the performance by clone detectors in detecting cloned co-change candidates. We have also increased the previous study's generalizability by adding two more software systems as subject systems and three more clone detection tools with four different configurations of CloneWorks (Type-1, Type-2 blind, Type-3 pattern, and Type-3 token), totaling eight subject systems and 12 clone detector executions. Therefore, our implementation has been upgraded from 6X6 to 12X8 (Clone detector X Subject Systems) in the study's current version. In this study, we have shown that the performance of clone detection tools in detecting cloned co-change fragments depends on some specific factors of clone detectors such as (i) the number of discovered clone fragments, (ii) the number of unique lines covered by those clone fragments, (iii) source file processing techniques, and (iv) type of detected clones. All the source code files, datasets, and processed results related to this study are publicly available \cite{cochangeByClones} for other researchers and practitioners to facilitate the continuation and reproducibility of this study. 

\section{Related Work}
\label{the-related-works}
Several studies \cite{Roy09comparisonand, jeff-evaluating, 4288192Comparison, ScenarioBasedComparison} have focused on ranking different clones detection tools based on their performance and accuracy in detecting different types of clone fragments. \citet{BaileyBurdComparison}  has compared three clones and two plagiarism detecting tools based on their performance in detecting cloned code in a single file or across different files. \citet{4288192Comparison} made a framework for comparing the performance of different clone detection tools from eight large C and Java programs having the size of source code almost 850 KLOC. One of the authors of this study also manually validated the dataset used in this study.  \citet{EvaluateRefactoring} compared three representative clone detection techniques from a refactoring perspective. Their criteria for comparison were the portability, kinds of clone reported, scalability, number of false positive, and number of useless clone detection from the results of those clone detection techniques. \citet{jeff-evaluating} reported ConQAT, iClones, NiCad, and SimCAD as excellent tools for detecting clones of all the three types (Type-1, Type-2, Type-3) based on their evaluation of eleven modern clone detection tools using four benchmark frameworks. \citet{Roy09comparisonand} did a qualitative comparison and evaluation of the latest clone detection approaches and tools, and made a benchmark called BigCloneBench \cite{BigCloneBenchCKRoyJRCordy} BigCloneBench included eight million manually verified clone pairs in a large inter-project source code dataset where the number of projects is larger than 25,000 and lines of code are above 365 million. They classify, compare, and evaluate different clone detectors based on the following point of view, (i) how the set of attributes in the different code fragments are overlapped, and (ii) what are the scenarios of creating Type-1, Type-2, Type-3, and Type-4 clones.  They also explained the procedure of using the result of their study for selecting the appropriate clone detectors in the context of particular application areas and restrictions. 

Besides proposing new clone detection mechanisms, some studies also compared their tools with a few existing tools. \citet{astDetectionComparisonBellon} utilizes suffix trees in abstract syntax trees to detect code clones and compared their technique with other few techniques using the Bellon benchmark for clone detectors \cite{DucasseStringMatchingCloneBallon}. Two other studies \cite{DucasseStringMatchingCloneBallon, CloneIntermediateRepresentationBallon} also measured the performance of their proposed clone detectors (based on string comparison and intermediate source transformation) utilizing Bellon’s framework. 

\citet{CloneIntermediateRepresentationBallon} showed their tool provides improved recall (with a slight drop in precision) compared to the source based clone detectors, and it also detects Type-3 clones. They also utilized Bellon’s corpus in their study and compared their technique with other standalone string and token-based clone detectors where they found little higher precision. All the studies that compared different clone detectors focused on the precision, recall, computational complexity, and memory used or detecting a specific type of clone fragments such as Type-1, Type-2, Type-3, or Type-4 during the detection approach of duplicated code in a codebase. 

Our study to compare clone detectors is entirely different from the previous comparisons. We do not want to compare clone detection tools based on the capability to detect clones. Our point of interest is to detect cloned co-change candidates during the software commit operations. \citet{Mondal-2014-PRC-2597073-2597104rankingCoChange} used the detected clone results of Nicad to predict and rank both the cloned and dissimilar co-change candidates (CCC and DCC) by analyzing evolutionary coupling from previously made change history. However, no other clone detection tool is included in their study to compare the performance of different clone detection tools in their prediction and ranking technique. We extended our previous study \cite{nadim-iwsc-2020} to compare the performance of 12 implementations of nine different clone detection tools based on the performance of detecting cloned co-change candidates using eight software systems written in C and Java programming languages.  We found no other study which has performed a similar comparison of clone detectors. We believe this investigation is the first to compare clone detection tools' in the perspective of specific software maintenance acclivity (doing change impact analysis by detecting cloned co-change candidates). 

\section{Methodology}
\label{the-methodology}
We have used eight open-source software systems, having varieties of size and application domains as subject systems in this study. The list of subject systems are in Table \ref{tab:subject-systems-cc}. To detect cloned co-change candidates from those subject systems, we executed 12 clone detection tools (Table \ref{tab:tool-configurations}) and analyzed obtained results to evaluate the performance of those clone detection tools. Our analysis evaluates these clone detection tools' performance in successfully suggesting cloned co-changed candidates (CCC) during the software evolution and determines the ranking of these tools based on this performance evaluation. To complete this study, we performed following set of operations. 

\begin{table}[htbp] 
\caption{\label{tab:subject-systems-cc}\textsc{Subject Systems}}
\centering
\begin{tabular}{|c|c|l|c|}
\hline
\textbf{Systems} & \textbf{Language} & \multicolumn{1}{c|}{\textbf{Domains}} & \textbf{Revisions} \\ \hline \hline
Brlcad           & C              & Computer Aided Design                 & 2115               \\ \hline
Camellia         & C              & Batch Job Server                      & 301                \\ \hline
Carol            & Java           & Game                                  & 1700               \\ \hline
Ctags            & C              & Code Def. Generator                   & 774                \\ \hline
Freecol          & Java           & Game                                  & 1950               \\ \hline
Jabref           & Java           & Reference Manager                     & 1545               \\ \hline
jEdit            & Java           & Text Editor                           & 4000               \\ \hline
Qmailadmin (QMA)       & C              & Mail System Manager                   & 317                \\ \hline
\end{tabular}
\end{table}

\subsection{Selecting the Subject Systems:} 
To select the subject systems, we first consider the popularity of programming languages and the availability of substantial revisions as essential factors. For example, the TIOBE Programming Community index \cite{TIOBE2021} (an indicator of the popularity of programming languages) recorded C, Python, Java, and C++ as the top-4 most popular programming languages in October 2021. We also consider the availability of substantial revisions and diversity in size and application domain as an essential factor for subject systems to produce a generalizable investigation result for this study. Besides these, we also reviewed some related studies \cite{Mondal:Association:Rules, Mondal:Co-change-recommendation, Mondal:Connectivity:co-changed, Mondal-2014-PRC-2597073-2597104rankingCoChange} and the programming language used in these subject systems. Considering all of these factors, we have selected the subject systems used in this study. Table \ref{tab:subject-systems-cc} includes the list of our eight subject systems having diverse sizes and application domains. Four of them are written in the C programming language, and the other four are in Java. As the C++ programming language share similar source code architecture, we believe the results obtained by subject systems in C also be applicable to the subject systems written in C++. Although the popularity index of the Python programming language is also very high, it does not share similar source code architecture (e.g., prior declaration of variable types, ending statements with a semicolon, use of brace to enclose statement block, etc.) with two other programming languages of this study.  Python is usually considered a scripting programming language. The primary purpose of using scripting languages (e.g., Python, Perl, Ruby, and PHP) is to develop software processes and technologies for the speedy production and coordination support for increasingly expanding web applications and web services \cite{ScriptingVSProgramming}. These scenarios motivate us to include subject systems written in C and Java programming languages in this study as they share homogeneous coding structure. However, we plan to extend this study, including a few more subject systems written in scripting languages such as Python, Perl, Ruby, and PHP in the future. 

\subsection{Selecting the Clone Detectors:} Considering some related studies about the ability of clone detection tools in detecting clone fragments, we have selected some promising mechanisms for this study, which can identify all the Type-1, Type-2, and Type-3 clones. We have taken CloneWorks \cite{CloneWorks-Jeff} as it is considered a fast and flexible clone detector for large-scale near-miss clone detection experiments. CloneWorks tool provides the ability to change its processing mechanism by changing its configuration files. We applied four different configurations of CloneWorks to detect Type-3 Pattern, Type-3 Token, Type-2 Blind, and Type-1 clones for investigating the impact of the types of clones in detecting cloned co-change candidates. We included Duplo \cite{DuploCloneDetection} as another Type-1 clone detector for comparing with Type-1 clones of CloneWorks in this investigation.  ConQAT \cite{conqat-clone-detecion}, iClones \cite{4812755iclones}, NiCad \cite{5970189Nicad}, and SimCAD \cite{6613857Simcad} have been reported as very good tools for detecting all types of clones in the study of \citet{jeff-evaluating}. Besides these, CCFinder \cite{CCFinderCloneDetection}, Deckard \cite{4222572Deckard}, iClones and NiCad are often considered as common examples of modern clone detectors that support Type-3 clone detection. CCFinder is known as a multi-linguistic token-based code clone detection system for large scale source code. The inclusion of CCFinder enriched the variation of detected clone fragments in the extended study.  To make more comparison of the performance of Type-1 clones in detecting cloned co-change candidates, we added Duplo in our study. We included Simian \cite{simianlink} in our analysis because of its ability to find cloned codes by line-by-line textual comparison supporting identifier renaming with a fast detection speed on the large repository and widespread use in several related studies \cite{simian-used-1, Wang-2013-SBC-2491411-2491420, impact-clone-maintenance, Cheung-clones-javascript, cloning-gnome-project}. NiCad, SimCAD, and Simian extract cloned code fragments from a software system's codebase using textual similarity among different code pieces. Deckard makes vector representation of different code fragments and then utilizes tree representation of those vectors to calculate the similarity among code pieces.  CCFinder, ConQAT, and iClones extract tokens from the source code and then use those tokens to find similar fragments.

\begin{table}[htbp]
\centering
\caption{\label{tab:tool-configurations}\textsc{Clone Detection Tools and Configurations}}
\addtolength{\tabcolsep}{-4.25pt}
\begin{tabular}{ll}
\hline
\multicolumn{1}{|l|}{\textbf{Tool}} & \multicolumn{1}{l|}{\textbf{Configuration}}                                                                                      \\ \hline \hline
\multicolumn{1}{|l|}{CCFinder}       & \multicolumn{1}{l|}{min. size: 50 tokens, min. token types: 12}                                                                                        \\ \hline
\multicolumn{1}{|l|}{\begin{tabular}[c]{@{}l@{}}CloneWorks Type-1\\(CLW-T1)\end{tabular}} &
\multicolumn{1}{l|}{termsplit=token, termproc=Joiner}                                                                                                  \\ \hline
\multicolumn{1}{|l|}{\begin{tabular}[c]{@{}l@{}}CloneWorks Type-2, Blind\\Renaming (CLW-T2B)\end{tabular}} &
\multicolumn{1}{l|}{\begin{tabular}[c]{@{}l@{}}cfproc=rename-blind,   cfproc=abstract literal, \\ termsplit=token, termproc=Joiner\end{tabular}}       \\ \hline
\multicolumn{1}{|l|}{\begin{tabular}[c]{@{}l@{}}CloneWorks Type-3,\\Pattern (CLW-T3P)\end{tabular}} &
\multicolumn{1}{l|}{\begin{tabular}[c]{@{}l@{}}cfproc=rename-blind, cfproc=abstract literal,\\ termsplit=line\end{tabular}}                            \\ \hline
\multicolumn{1}{|l|}{\begin{tabular}[c]{@{}l@{}}CloneWorks Type-3,\\Token (CLW-T3T)\end{tabular}} &
\multicolumn{1}{l|}{\begin{tabular}[c]{@{}l@{}}termsplit=token,   termproc=FilterOperators, \\ termproc=FilterSeperators\end{tabular}}                 \\ \hline
\multicolumn{1}{|l|}{ConQAT}         & \multicolumn{1}{l|}{block clones, clone   min-length=5, gap ratio=0.3}                                                                                 \\ \hline
\multicolumn{1}{|l|}{Deckard}        & \multicolumn{1}{l|}{\begin{tabular}[c]{@{}l@{}}min. size: 30   tokens, 5 token stride, \\ min. 85\% similarity\end{tabular}}                           \\ \hline
\multicolumn{1}{|l|}{Duplo}          & \multicolumn{1}{l|}{min. size: 10 lines, min. characters/line:1}                                                                                       \\ \hline
\multicolumn{1}{|l|}{iClones}        & \multicolumn{1}{l|}{\begin{tabular}[c]{@{}l@{}}minimum block: 30, minimum   clone: 50, \\ All Transformation\end{tabular}}                             \\ \hline
\multicolumn{1}{|l|}{Nicad}          & \multicolumn{1}{l|}{\begin{tabular}[c]{@{}l@{}}block clones,   blind renaming, max. threshold=0.3,\\ minimum lines=5, maximum lines=2500\end{tabular}} \\ \hline
\multicolumn{1}{|l|}{SimCAD}         & \multicolumn{1}{l|}{block clones,   Source Transformation= generous}                                                                                   \\ \hline
\multicolumn{1}{|l|}{Simian}         & \multicolumn{1}{l|}{min. size: 5   lines, normalize literals/identifiers}                                                                              \\ \hline
\end{tabular}
\end{table}

\subsection{Selecting Identical Parameter Configurations of the Clone Detectors} 
Taking an identical configuration of different clone detectors while applying them on different subject systems to identify cloned code fragments is essential. We compared them with each other based on their capability to successfully suggest cloned co-change candidates. Identified clones from each clone detectors have a crucial impact on their performance in suggesting cloned co-change candidates. \citet{Wang-2013-SBC-2491411-2491420} reported taking different configurations of clone detectors may change the result of detected clones, and the result can be very good or terrible depending on the taken configuration. This scenario is known as a confounding configuration choice problem, and it should be handled very carefully while determining the configuration to be taken to detect cloned fragments using any clone detection tool. 
Our configuration of different tools is shown in Table \ref{tab:tool-configurations}. As our goal is to compare different tools with each other, configuring them similarly while clone detection will provide consistent results for a fair comparison. We have also tried to have identical configuration values to \citet{jeff-evaluating}, which they conducted to compare different clone detectors based on their efficiency in detecting cloned fragments. We provided a 70\% similarity threshold for all the clone detection tools (except Deckard), which takes similarity dissimilarity value as a parameter. We have used 85\% as the similarity threshold for Deckard because, in an initial manual inspection of the detected results, we found a massive number of clone fragments in Deckard's results compared to the other clone detectors while using the similarity threshold 70\% like the others. We also identified many clone classes having the same fragments (self duplicate fragments) as the clone fragments of that class. To minimize those duplicated fragments, we tried other similarity threshold values such as 75\%, 80\%, and  85\%. We found that increasing the similarity threshold reduces the self duplicated fragments, and we took the detected clone results applying the 85\% similarity threshold for Deckard in all the subject systems. \citet{jeff-evaluating} also used 85\% similarity while running Deckard for Mutation Framework. We have also selected identical parameter values such as the minimum number of tokens, the minimum number of lines for different clone detection tools. As we wanted to compare different clone detectors based on their capability of successfully suggesting cloned co-change candidates, it was essential to configure them identically during detecting clones.

\subsection{Identifying Cloned and Dissimilar Co-change Candidates}
We prepared the cloned co-change candidates by processing all the revisions of all the subject systems, which we mention in Table \ref{tab:subject-systems-cc}. We show the summary of revisions processed from each of the subject systems in Table \ref{tab:data-summary}. Initially, we downloaded all the source files of all the subject systems' revisions from their respective SVN repositories. We then applied the \textbf{Unix diff} operation between all the files of each revision with the corresponding files in the next revision to extract changes in these revisions. We prefer the Unix diff over git-diff, as there are four different variants (e.g., Myers, Minimal, Patience, and Histogram) of git-diff \footnote{https://git-scm.com/docs/git-diff}, and \citet{DifferentDiff} show that different git-diff variants can provide different source code lines as the results of git-diff using the same set of input source files. Besides these, instead of working directly on the mirrored GitHub repositories of the subject systems, we extracted all the source code files of each revision from each subject system to apply the clone detection tools to find the cloned fragments in these revisions. The clone fragments from the results of clone detectors are compared with the results of Unix diff to produce the ground truth of cloned co-change candidates for our investigation and evaluate the performance of clone detectors for detecting cloned co-change candidates. Therefore, we applied the Unix diff directly to each source code file pair from which clone detectors detected the clone fragments. We use the Unix diff command instead of any variant of four available git-diff commands to make the source code fragments consistent in the line numbers in both the output of clone detectors and the Unix diff.   

We identify some specific change information such as (i) the name of the files which are changed, (ii) the beginning line numbers of each of the changes, and (iii) the ending line numbers of the corresponding changes. We repeat the change extraction process for each revision (excluding the last one) in all the subject systems. Suppose we find n changed fragments in a software revision. Then, for every single change fragment, we consider the remaining n-1 fragments as the combination of cloned and dissimilar co-change candidates for that revision. In the following step, we describe the method for excluding the dissimilar co-change fragments to produce the ground truth of cloned co-change candidates. 

\begin{table}[htbp]
\centering
\caption{\label{tab:data-summary}\textsc{Revisions Processed from each of the Subject Systems}}
\small\addtolength{\tabcolsep}{-3.0pt}
\begin{tabular}{|l|c|c|c|c|c|c|c|c|}
\hline

\textbf{\begin{tabular}[c]{@{}l@{}}Revision\end{tabular}} & \textbf{Brlcad} & \textbf{Camellia} & \textbf{Carol} & \textbf{Ctags} & \textbf{Freecol} & \textbf{Jabref} & \textbf{jEdit} & \textbf{QMA} \\ \hline \hline
Any change                                                                   & 2113            & 301               & 1700           & 774            & 1001             & 1540            & 215            & 317                                                            \\ \hline
\begin{tabular}[c]{@{}l@{}}$\geq$ 1 change \\ to C or Java\end{tabular}                                                         & 660             & 163               & 454            & 447            & 836              & 860             & 145            & 35                                                             \\ \hline
\begin{tabular}[c]{@{}l@{}}$>$ 1 change \\ to C or Java\end{tabular}  & 553             & 155               & 430            & 330            & 833              & 755             & 145            & 25                                                             \\ \hline
\end{tabular}
\end{table}

\subsection{Identifying Cloned Co-change Candidates (Ground Truth)} 
Even though the changes extracted from two adjacent revisions (e.g., revision n and n+1) are co-change candidates of each other, they might not be a set of cloned co-change candidates. Thus, for example, a group of co-change candidates might contain some dissimilar change candidates who changed independently, and they might not be a clone of any other changes of that co-change group. The inclusion of such non-cloned changes into our calculation can drop the detection accuracy of clone detectors. Therefore, we utilize the results from all the 12 clone detectors used in this study to produce the ground truth of cloned co-change candidates before calculating the performance metrics (precision, recall, and f1~scores) for each clone detector. We first considered each code fragment's change as target change and tried to predict all the other changes in the same revision (e.g., co-changes) utilizing each of the 12 clone detection techniques. We find some change fragments where none of the 12 clone detection techniques can detect any cloned co-change candidate. Therefore, we consider those change fragments as non-cloned (dissimilar) co-change fragments for the set of clone detectors used in this study. Thus, we exclude the identified non-cloned co-change candidates to prepare the ground truth of cloned co-change candidates for comparing the 12 clone detection techniques.  Including only the dependent (cloned) co-changes while calculating the performance metrics (precision, recall, and f1~scores) is necessary to make a fair comparison among the clone detection tools. Finally, we compare all the tools based on their calculated performance metrics in detecting cloned co-change candidates.

All the cloned fragments might not be cloned co-change candidates. For example, a clone class might contain a large number of clone fragments. However, whenever we attempt to change a code fragment, we can see that only a few clone fragments could change together with that particular fragment, and in such a case, those few other fragments are the actual cloned co-change candidates. The remaining clone fragments are not the cloned co-change candidates in such cases. Therefore, our ground truth does not directly depend on clone detection tools result. We extracted the co-change history using the Unix diff operation from the code repositories of the subject systems. We use this co-change history to make our ground truth. For example, let us assume that five code fragments changed together in a particular commit operation. Then, we consider the remaining four as the true (combination of cloned and dissimilar) co-change candidates for each of these five code fragments. Then we took the union of detected cloned co-change candidates using all the 12 clone detectors to produce the final ground truth for comparing the tools. As a gold data set of co-change candidates currently does not exist, we believe our considered ground truth is reasonable for our comparison scenario. 

\begin{figure}
\centering
\includegraphics[width=\columnwidth] {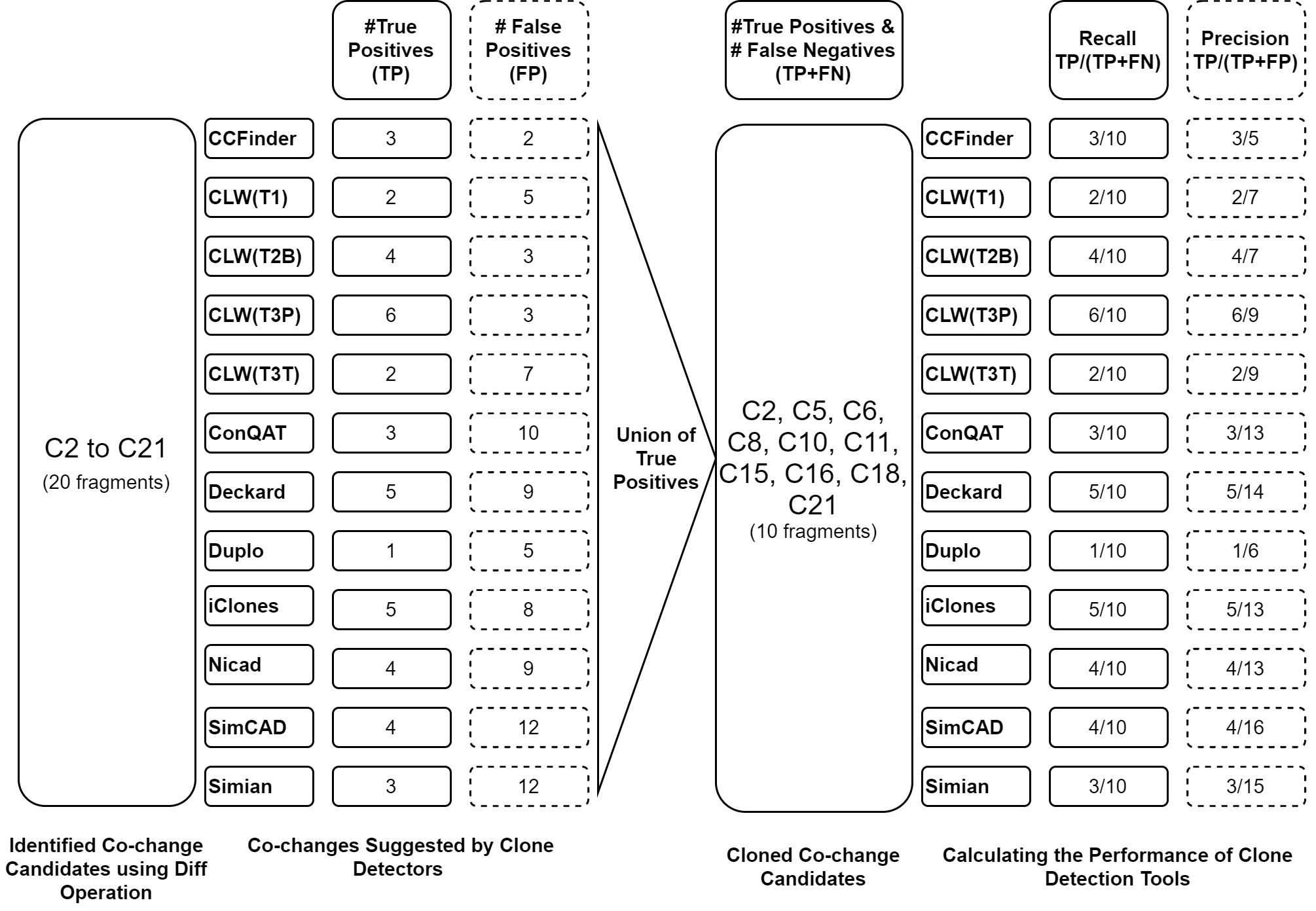}
\caption{This diagram explains how we calculated the Recall and Precision in one revision. We first identified all the cloned and dissimilar co-change candidates as ground truth using the Unix diff operation, then we have used the results of each of the clone detection tools to identify those cloned co-change candidates. True Positives (TP) are the co-changes that we successfully identified using the clone detectors. Our process also provided some False Positives (FP) in each revision. To separate the cloned co-change candidates from all the previously identified co-change candidates, we took the union of all the TPs from all the clone detection tools. The Recall is the ratio of TPs and the number of cloned co-change candidates (TP+FN), and the Precision is the ratio of TPs and the number of total co-change suggestions (TP+FP). We repeated the process for all the revisions of all the subject systems used in our investigation.}
\label{fig:CalculatingCC}
\end{figure}

\subsection{Demonstrating a Running Example} 
Figure \ref{fig:CalculatingCC} provides a demonstration of a running example in calculating each of the clone detection tools' performance metric values. Let us consider that 21 changes, C1 to C21, occurred in the codebase during a particular commit operation in a subject system. We extract these changes utilizing the UNIX diff operation. If we consider an arbitrary change (e.g., C1) as the target change, the other 20 changes (e.g., C2 to C21) will be considered the co-change candidates (combination of both cloned and dissimilar co-change candidates) of C1.  We apply different clone detection tools to detect these co-change candidates considering C1 as the target change. To do so, we first take the clone class from the results of the clone detection tool and find if any code fragment from that clone class intersects with C1. Here, the intersection of any two code fragments means they share at least one common line. These code fragments could be fully overlapped or partially overlapped to consider them as intersecting code fragments. Therefore, both the fully and partially overlapped code fragments would be considered as intersecting code fragments. We consider the intersecting code fragment from the clone class as the equivalent code fragment of the target change C1. Therefore, the other code fragments in that intersecting clone class should be co-change with the target change C1. There code fragments from the clone class are the predicted co-change candidates of C1.  

\vspace{0.35cm}
Let us assume, analyzing the detected clone fragments of one of the clone detection tools, Deckard \cite{4222572Deckard}, we find 14 code fragments as clone candidates of C1, among those 14 fragments, five intersects with the five co-change candidates of C1 (e.g., C2, C6, C8, C15, C21) from those 20 fragments. Thus, we will consider those five fragments as true positives (TP) of cloned co-change candidates (CCC), and $14-5=9$ will be the number of false positive (FP) value. Similarly, using the clone fragments of NiCad, if we detect 13 code fragments as clone candidates of C1, and four of them intersects with four of co-change candidates (e.g., C5, C10, C16, C18), then we find $TP=4$, and $FP=13-4=9$. We continued the same approach for all the clone detection tools to detect co-change candidates of C1. If we take the union of all the true positives obtained from the results of all the clone detectors, we get ten unique change fragments (C2, C5, C8, C15, C21, C5, C8, C10, C15, C21). We will then consider those ten unique change fragments as cloned co-change candidates and this will be used as the number of TP+FN value while calculating the recall of clone detection tools for detecting cloned co-change of C1.  Using the number of TP and FP values, we can also calculate the precision of clone detection tools for C1. We finally calculated average recall, average precision in each of the subject systems for all the clone detection tools. We compare the clone detection tools based on the F1~Scores calculated by using the average recall and precision shown in the equation \ref{eq-f1-score}. 


\subsection{Calculating the Evaluation Metrics}
Let us assume, while examining a specific commit operation, we found the number of fragments changed with this commit is $n$. In each step, we consider one of those fragments as the target fragment and the remaining $n-1$ fragments as the co-changed candidates for that target fragment. Among the $n-1$ fragments, there could be some non-cloned fragments that might have changed independently. We exclude such non-cloned fragments (demonstrated in Figure \ref{fig:CalculatingCC}) by taking the union of detected cloned co-change candidates by all the clone detection tools used in this study. After excluding non-cloned fragments, we get the \textbf{Cloned Co-changed Candidate} (CCC) for each of the target fragments. Now, we find the cloned fragment from the results of clone detectors intersecting with the target fragment. The other fragments in the intersecting clone fragment class are considered the \textbf{Predicted Cloned Co-change} (PCC) candidates. We now determine how many of these PCC intersect (shares common line number) with the CCC to obtain the number of detected cloned co-change candidates by the clone detector. 
\begin{equation}
    \text{Recall} = \frac{|PCC \cap CCC|}{|CCC|}
\end{equation}

\begin{equation}
    \text{Precision} = \frac{|PCC \cap CCC|}{|PCC|}
\end{equation}

\begin{equation}
\label{eq-f1-score}
    F1~Score = \frac{2 \times Precision \times Recall}{\text{Precision} + Recall}
\end{equation}

These counts of predicted and actually co-changed candidates are considered as the \textbf{true positives} to calculate Recall, Precision, and F1~Score. We calculate these using the following equations (Eq. 1, 2, and 3). 

We repeat the calculating process of Recall and Precision for all the changes in each of the subject systems with the detected clone fragments generated by all the clone detection tools. We then calculate the F1~Score of the clone detectors for each of the subject systems by taking the average values of Recall and Precision, which is reported in Table \ref{tab:detection-f1-score}. We reported the ranking of the tools considering individual ranks in each of the subject systems in Table \ref{tab:final-ranking-sum-of-ranks}.

\subsection{Producing the Final Rank List:} 
We produce the final rank list based on the ranks of the clone detectors in each of the subject systems. To produced the final rank list of 12 clone detection techniques, we considered their performance in all the eight individual subject systems.  Our ranking approach is demonstrated in Table \ref{tab:final-ranking-sum-of-ranks}, which shows both the final and individual ranks of clone detectors. The columns $S1$ to $S8$ indicates individual ranks of clone detectors in each subject system. Individual ranks are measured based on the F1~scores of each clone detector shown in Table \ref{tab:detection-f1-score}. The highest F1~Score in Table \ref{tab:detection-f1-score} got rank-1, and similarly the lowest one got rank-12 in the respective position of Table \ref{tab:final-ranking-sum-of-ranks}. Thus, every clone detection technique has eight rank values (smaller value represents the better performance), obtained in the eight software systems. The right-most column (\textbf{Rank}) shows the final rank of the clone detectors considering all the individual ranks. As smaller rank values indicate better performance of the clone detectors in detecting cloned co-change candidates, a clone detector that performed well for more subject systems, will obtain smaller $\sum\limits_{i=1}^8 S_i$, and received a higher position in the final rank table. 

\section{Experimental Result}
\label{the-experimental-result}
We investigate  12 clone detection techniques on the thousands of revisions from eight C and Java programming language-based software systems to detect cloned co-change candidates. In this section, we describe the obtained results and answer the research questions based on this investigation.  
We produce the final rank list of all the clone detection techniques in Table \ref{tab:final-ranking-sum-of-ranks}. Considering the final rank list, we find:
\begin{enumerate*}[label=(\roman*)]
  \item the tools which are good in detecting all types (1/2/3) of clones are also good in detecting cloned co-change candidates. 
  \item the top two tools in the final rank list are the Type-3 configurations of CloneWorks (prior to detecting clone fragments one configuration first splits the source files into lines while the other configuration splits source files into tokens), and the next two clone detectors that perform well are Deckard and CCFinder. Therefore, to detect cloned co-change candidates, those tools (all are pattern and token-based) are the best choices compared to the other tools compared in this study.
  \item From this ranking, we also find that text-based clone detectors (such as Duplo or CloneWorks Type-1) are not suitable for detecting cloned co-change candidates.
  \item Our comparison in Figure \ref{fig:AverageLineCoveredPerSS} also shows that the clone detectors which detect a higher number of clone fragments and cover a higher number of unique lines in the source files perform well when detecting cloned co-change candidates. 
\end{enumerate*}
We performed the Wilcoxon Signed-Rank Test \cite{wilcoxon-signed-rank-test, wilcoxon-signed-rank-test-rosner} to verify whether the F1~Scores in all of the eight subject systems of the tools which got higher ranks in the final rank list (Table \ref{tab:final-ranking-sum-of-ranks}) are significantly better compared to the other clone detection tools or not. The results of our significance test are described in Section \ref{sec-wilcoxon-singed-rank-test}. A summary of our significance test results is in Table \ref{tab:cochange-wilcoxon-rank-test}, which shows that four out of the 12 clone detection techniques of this study perform significantly better than the other techniques in detecting cloned co-change candidates.

\begin{table}[htbp]
\centering
\addtolength{\tabcolsep}{-4pt}
\caption{\textsc{F1~Score of Different Tools in Detecting Cloned Co-change}}
\label{tab:detection-f1-score}
\begin{tabular}{|l|c|c|c|c|c|c|c|c|}
\hline
\multicolumn{1}{|c|}{\textbf{Tool}} & \textbf{Brlcad} & \textbf{Camellia} & \textbf{Carol} & \textbf{Ctags} & \textbf{Freecol} & \textbf{Jabref} & \textbf{jEdit} & \textbf{QMA} \\ \hline \hline
CCFinder                                & 0.30            & 0.23              & 0.20           & 0.16           & 0.09             & 0.15            & 0.07           & 0.16         \\ \hline
CLW-T1                                 & 0.09            & 0.04              & 0.06           & 0.03           & 0.03             & 0.05            & 0.04           & 0.11         \\ \hline
CLW-T2B                            & 0.13            & 0.07              & 0.22           & 0.12           & 0.08             & 0.13            & 0.08           & 0.17         \\ \hline
CLW-T3P                          & 0.32            & 0.16              & 0.36           & 0.25           & 0.15             & 0.30            & 0.35           & 0.49         \\ \hline
CLW-T3T                            & 0.27            & 0.18              & 0.29           & 0.24           & 0.11             & 0.20            & 0.20           & 0.42         \\ \hline
ConQAT                                  & 0.28            & 0.10              & 0.12           & 0.15           & 0.08             & 0.12            & 0.08           & 0.08         \\ \hline
Deckard                                 & 0.19            & 0.57              & 0.21           & 0.15           & 0.30             & 0.14            & 0.18           & 0.41         \\ \hline
Duplo                                   & 0.12            & 0.01              & 0.03           & 0.03           & 0.01             & 0.02            & 0.00           & 0.00         \\ \hline
iClones                                 & 0.26            & 0.15              & 0.08           & 0.08           & 0.03             & 0.09            & 0.05           & 0.10         \\ \hline
Nicad                                   & 0.12            & 0.06              & 0.16           & 0.21           & 0.04             & 0.10            & 0.12           & 0.03         \\ \hline
SimCAD                                  & 0.17            & 0.07              & 0.15           & 0.04           & 0.05             & 0.10            & 0.06           & 0.10         \\ \hline
Simian                                  & 0.25            & 0.16              & 0.06           & 0.11           & 0.03             & 0.07            & 0.03           & 0.06         \\ \hline
\end{tabular}
\end{table}

\begin{table}[htbp]
\centering
\caption{\label{tab:summary-atc-acc}\textsc{Actual Target Changes (ATC) and Cloned Co-changed Candidates (CCC)}}
\begin{tabular}{lcccc}
\hline
\multicolumn{1}{|c|}{\textbf{System}}    & \multicolumn{1}{c|}{\textbf{\# ATC}} & \multicolumn{1}{c|}{\textbf{\# CCC}}  & \multicolumn{1}{c|}{\textbf{\% ATC}} & \multicolumn{1}{c|}{\textbf{\% CCC}} \\ \hline \hline
\multicolumn{1}{|l|}{Brlcad}         & \multicolumn{1}{c|}{2,909}            & \multicolumn{1}{c|}{33,578}            & \multicolumn{1}{c|}{7.45}            & \multicolumn{1}{c|}{1.89}            \\ \hline
\multicolumn{1}{|l|}{Camellia}       & \multicolumn{1}{c|}{8,052}            & \multicolumn{1}{c|}{346,140}           & \multicolumn{1}{c|}{20.61}           & \multicolumn{1}{c|}{19.46}           \\ \hline
\multicolumn{1}{|l|}{Carol}          & \multicolumn{1}{c|}{4,582}            & \multicolumn{1}{c|}{254,311}           & \multicolumn{1}{c|}{11.73}           & \multicolumn{1}{c|}{14.29}           \\ \hline
\multicolumn{1}{|l|}{Ctags}          & \multicolumn{1}{c|}{718}             & \multicolumn{1}{c|}{3,648}             & \multicolumn{1}{c|}{1.84}            & \multicolumn{1}{c|}{0.21}            \\ \hline
\multicolumn{1}{|l|}{Freecol}        & \multicolumn{1}{c|}{6,865}            & \multicolumn{1}{c|}{265,213}           & \multicolumn{1}{c|}{17.57}           & \multicolumn{1}{c|}{14.91}           \\ \hline
\multicolumn{1}{|l|}{Jabref}         & \multicolumn{1}{c|}{8,313}            & \multicolumn{1}{c|}{455,469}           & \multicolumn{1}{c|}{21.28}           & \multicolumn{1}{c|}{25.60}           \\ \hline
\multicolumn{1}{|l|}{jEdit}          & \multicolumn{1}{c|}{5,122}            & \multicolumn{1}{c|}{323,277}           & \multicolumn{1}{c|}{13.11}           & \multicolumn{1}{c|}{18.17}           \\ \hline
\multicolumn{1}{|l|}{QMA}            & \multicolumn{1}{c|}{2,508}            & \multicolumn{1}{c|}{97,396}            & \multicolumn{1}{c|}{6.42}            & \multicolumn{1}{c|}{5.47}            \\ \hline
\multicolumn{1}{|l|}{\textbf{Total}} & \multicolumn{1}{c|}{\textbf{39,069}}  & \multicolumn{1}{c|}{\textbf{1,779,032}} & \multicolumn{1}{c|}{\textbf{100}} & \multicolumn{1}{c|}{\textbf{100}} \\ \hline
\end{tabular}
\end{table}

\begin{table}[htbp]
\caption{\textsc{Final Rank of Tools considering individual subject system rankings}}
\label{tab:final-ranking-sum-of-ranks}
\centering
\begin{tabular}{lcccccccccc}
\hline
\multicolumn{1}{|c|}{\textbf{\begin{tabular}[c]{@{}c@{}}Tool\end{tabular}}} & \multicolumn{1}{c|}{\textbf{S1}} & \multicolumn{1}{c|}{\textbf{S2}} & \multicolumn{1}{c|}{\textbf{S3}} & \multicolumn{1}{c|}{\textbf{S4}} & \multicolumn{1}{c|}{\textbf{S5}} & \multicolumn{1}{c|}{\textbf{S6}} & \multicolumn{1}{c|}{\textbf{S7}} & 
\multicolumn{1}{c|}{\textbf{S8}} & \multicolumn{1}{c|}{\textbf{\begin{tabular}[c]{@{}c@{}}$\sum\limits_{i=1}^8 S_i$\end{tabular}}} & 
\multicolumn{1}{c|}{\textbf{\begin{tabular}[c]{@{}c@{}}Rank\end{tabular}}} \\ \hline \hline
\multicolumn{1}{|l|}{\textbf{CLW-T3P}}                                                   & \multicolumn{1}{c|}{1}           & \multicolumn{1}{c|}{4}           & \multicolumn{1}{c|}{1}           & \multicolumn{1}{c|}{1}           & \multicolumn{1}{c|}{2}           & \multicolumn{1}{c|}{1}           & \multicolumn{1}{c|}{1}           & \multicolumn{1}{c|}{1}           & \multicolumn{1}{c|}{12}                                                               & \multicolumn{1}{c|}{\textbf{1}}                                                    \\ \hline
\multicolumn{1}{|l|}{\textbf{CLW-T3T}}                                                     & \multicolumn{1}{c|}{4}           & \multicolumn{1}{c|}{3}           & \multicolumn{1}{c|}{2}           & \multicolumn{1}{c|}{2}           & \multicolumn{1}{c|}{3}           & \multicolumn{1}{c|}{2}           & \multicolumn{1}{c|}{2}           & \multicolumn{1}{c|}{2}           & \multicolumn{1}{c|}{20}                                                               & \multicolumn{1}{c|}{\textbf{2}}                                                    \\ \hline
\multicolumn{1}{|l|}{\textbf{Deckard}}                                                          & \multicolumn{1}{c|}{7}           & \multicolumn{1}{c|}{1}           & \multicolumn{1}{c|}{4}           & \multicolumn{1}{c|}{6}           & \multicolumn{1}{c|}{1}           & \multicolumn{1}{c|}{4}           & \multicolumn{1}{c|}{3}           & \multicolumn{1}{c|}{3}           & \multicolumn{1}{c|}{29}                                                               & \multicolumn{1}{c|}{\textbf{3}}                                                    \\ \hline
\multicolumn{1}{|l|}{\textbf{CCFinder}}                                                         & \multicolumn{1}{c|}{2}           & \multicolumn{1}{c|}{2}           & \multicolumn{1}{c|}{5}           & \multicolumn{1}{c|}{4}           & \multicolumn{1}{c|}{4}           & \multicolumn{1}{c|}{3}           & \multicolumn{1}{c|}{7}           & \multicolumn{1}{c|}{5}           & \multicolumn{1}{c|}{32}                                                               & \multicolumn{1}{c|}{\textbf{4}}                                                    \\ \hline
\multicolumn{1}{|l|}{\textbf{CLW-T2B}}                                                     & \multicolumn{1}{c|}{9}           & \multicolumn{1}{c|}{8}           & \multicolumn{1}{c|}{3}           & \multicolumn{1}{c|}{7}           & \multicolumn{1}{c|}{5}           & \multicolumn{1}{c|}{5}           & \multicolumn{1}{c|}{5}           & \multicolumn{1}{c|}{4}           & \multicolumn{1}{c|}{46}                                                               & \multicolumn{1}{c|}{\textbf{5}}                                                    \\ \hline
\multicolumn{1}{|l|}{\textbf{ConQAT}}                                                           & \multicolumn{1}{c|}{3}           & \multicolumn{1}{c|}{7}           & \multicolumn{1}{c|}{8}           & \multicolumn{1}{c|}{5}           & \multicolumn{1}{c|}{6}           & \multicolumn{1}{c|}{6}           & \multicolumn{1}{c|}{6}           & \multicolumn{1}{c|}{9}           & \multicolumn{1}{c|}{50}                                                               & \multicolumn{1}{c|}{\textbf{6}}                                                    \\ \hline
\multicolumn{1}{|l|}{\textbf{iClones}}                                                          & \multicolumn{1}{c|}{11}          & \multicolumn{1}{c|}{10}          & \multicolumn{1}{c|}{6}           & \multicolumn{1}{c|}{3}           & \multicolumn{1}{c|}{8}           & \multicolumn{1}{c|}{7}           & \multicolumn{1}{c|}{4}           & \multicolumn{1}{c|}{11}          & \multicolumn{1}{c|}{60}                                                               & \multicolumn{1}{c|}{\textbf{7}}                                                    \\ \hline
\multicolumn{1}{|l|}{\textbf{Simian}}                                                           & \multicolumn{1}{c|}{5}           & \multicolumn{1}{c|}{6}           & \multicolumn{1}{c|}{9}           & \multicolumn{1}{c|}{9}           & \multicolumn{1}{c|}{10}          & \multicolumn{1}{c|}{9}           & \multicolumn{1}{c|}{9}           & \multicolumn{1}{c|}{7}           & \multicolumn{1}{c|}{64}                                                               & \multicolumn{1}{c|}{\textbf{8}}                                                    \\ \hline
\multicolumn{1}{|l|}{\textbf{Nicad}}                                                            & \multicolumn{1}{c|}{8}           & \multicolumn{1}{c|}{9}           & \multicolumn{1}{c|}{7}           & \multicolumn{1}{c|}{10}          & \multicolumn{1}{c|}{7}           & \multicolumn{1}{c|}{8}           & \multicolumn{1}{c|}{8}           & \multicolumn{1}{c|}{8}           & \multicolumn{1}{c|}{65}                                                               & \multicolumn{1}{c|}{\textbf{9}}                                                    \\ \hline
\multicolumn{1}{|l|}{\textbf{SimCAD}}                                                           & \multicolumn{1}{c|}{6}           & \multicolumn{1}{c|}{5}           & \multicolumn{1}{c|}{11}          & \multicolumn{1}{c|}{8}           & \multicolumn{1}{c|}{11}          & \multicolumn{1}{c|}{10}          & \multicolumn{1}{c|}{11}          & \multicolumn{1}{c|}{10}          & \multicolumn{1}{c|}{72}                                                               & \multicolumn{1}{c|}{\textbf{10}}                                                   \\ \hline
\multicolumn{1}{|l|}{\textbf{CLW-T1}}                                                          & \multicolumn{1}{c|}{12}          & \multicolumn{1}{c|}{11}          & \multicolumn{1}{c|}{10}          & \multicolumn{1}{c|}{11}          & \multicolumn{1}{c|}{9}           & \multicolumn{1}{c|}{11}          & \multicolumn{1}{c|}{10}          & \multicolumn{1}{c|}{6}           & \multicolumn{1}{c|}{80}                                                               & \multicolumn{1}{c|}{\textbf{11}}                                                   \\ \hline
\multicolumn{1}{|l|}{\textbf{Duplo}}                                                            & \multicolumn{1}{c|}{10}          & \multicolumn{1}{c|}{12}          & \multicolumn{1}{c|}{12}          & \multicolumn{1}{c|}{12}          & \multicolumn{1}{c|}{12}          & \multicolumn{1}{c|}{12}          & \multicolumn{1}{c|}{12}          & \multicolumn{1}{c|}{12}          & \multicolumn{1}{c|}{94}                                                               & \multicolumn{1}{c|}{\textbf{12}}                                                   \\ \hline
\multicolumn{11}{l}{\textit{\begin{tabular}[c]{@{}l@{}}\textbf{* S1-S8} represents sequence of eight subject systems used in this study.\end{tabular}}}                                                                                                                                                                                                                                                                                                                                                                                                          
\end{tabular}
\end{table}

\subsection{Answer to the \textbf{RQ1}}
\textbf{How can we compare different clone detection tools based on the performance in detecting cloned co-change candidates?}

\begin{figure}
\centering
\includegraphics[width=\textwidth] {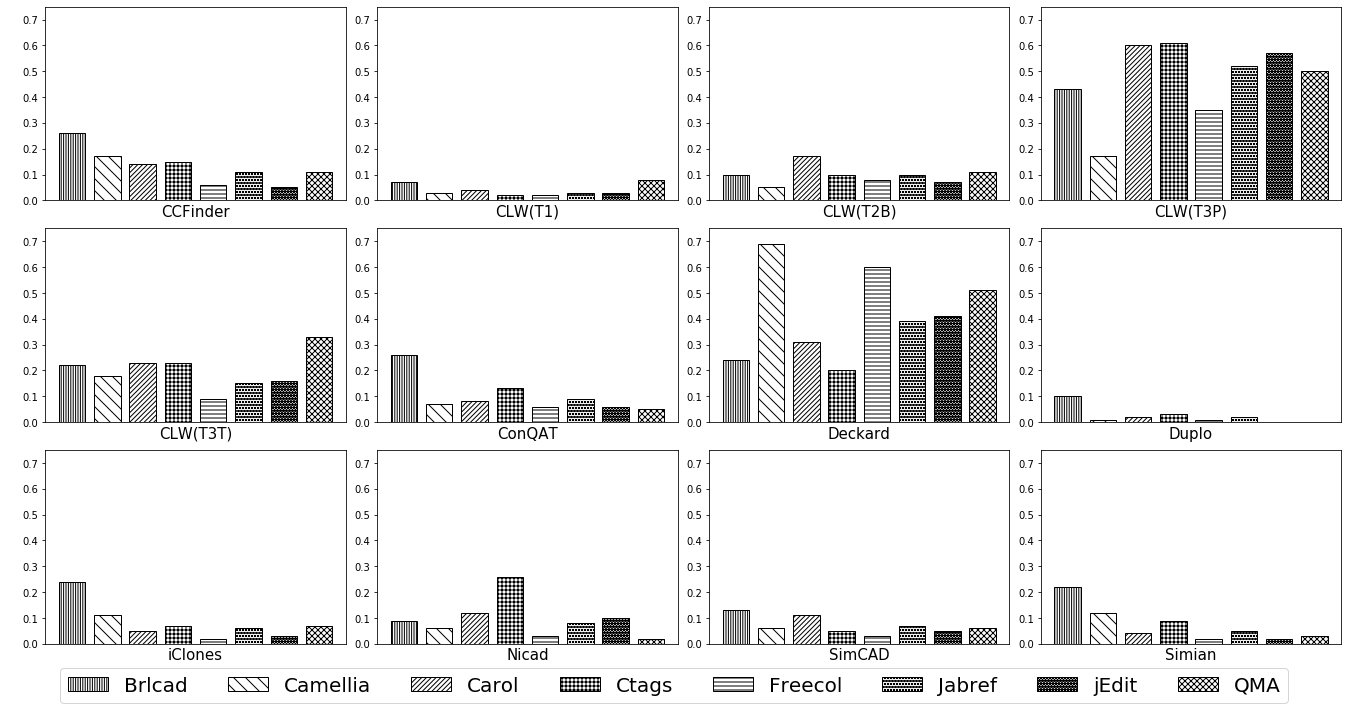}
\caption{Average recall of different tools}
\label{fig:jssRecall}
\end{figure}

\begin{figure}
\centering
\includegraphics[width=\textwidth] {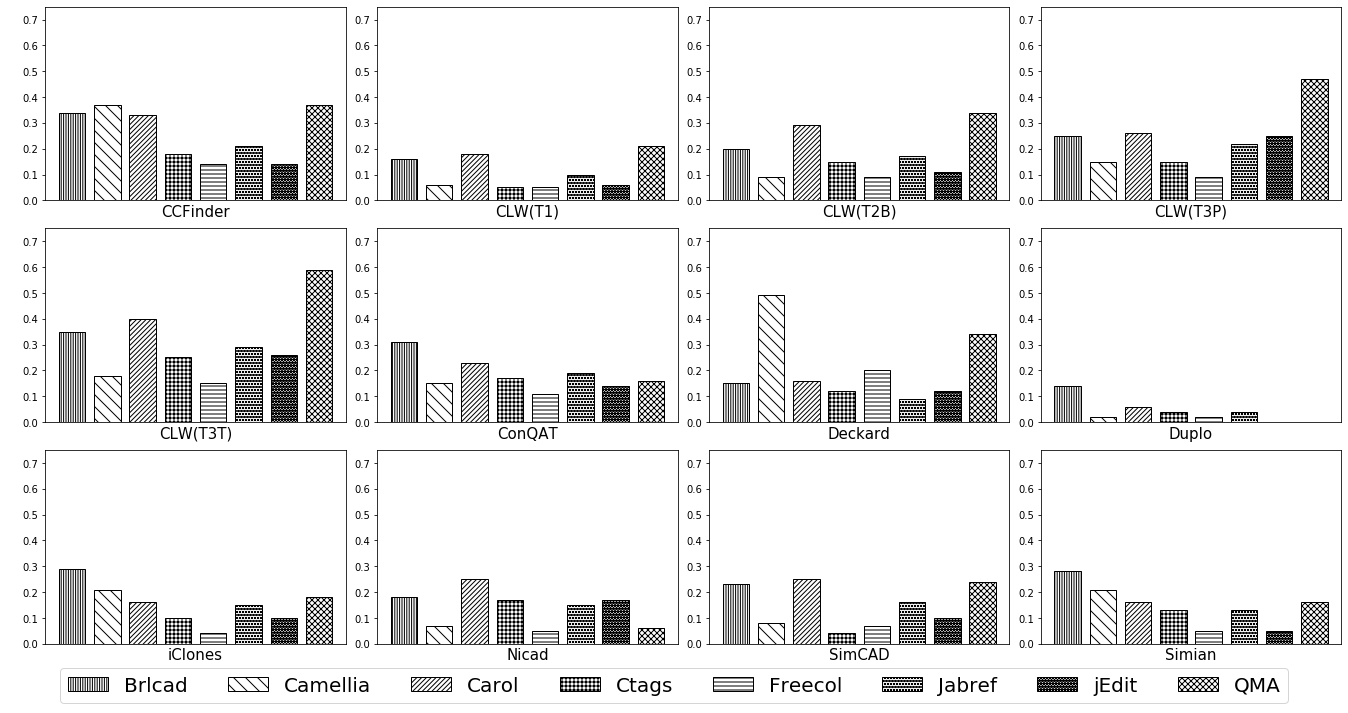}
\caption{Average precision of different tools}
\label{fig:jssPrecision}
\end{figure}

The key experimental results are in Figure \ref{fig:jssRecall}, Figure \ref{fig:jssPrecision}, Table \ref{tab:detection-f1-score}, Table \ref{tab:summary-atc-acc},  and Table \ref{tab:final-ranking-sum-of-ranks}, where Fig. \ref{fig:jssRecall} and \ref{fig:jssPrecision} show the average Recall and average Precision of each of the clone detection tools. Table \ref{tab:summary-atc-acc} shows the summary of target changes and detected cloned co-change candidates for those target changes in each of the subject systems.  We found the highest and lowest percentage of target change and its cloned co-change candidates from Jabref and Ctags. Table \ref{tab:detection-f1-score} shows the F1~Score of each of the clone detectors in each of the subject systems. The F1~Score is calculated using Equation (3). Our experimental results concluded in Table \ref{tab:final-ranking-sum-of-ranks}, which shows that CLW-T3P, CLW-T3T, and Deckard shows top performance (Rank 1 or 2 based on F1~Score) in most of the subject systems compared to all the other tools. The summary of the results in Table \ref{tab:final-ranking-sum-of-ranks} also shows that among the subject systems, CLW-T3P is the best in all the subject systems except Camellia and Freecol, where Deckard is showing the best cloned co-change detection performance. CLW-T3T shows the second-best performance in most of the subject systems. An overall observation on individual rankings of different clone detection techniques reveals that CLW-T3P, Deckard, CLW-T3T, CCFinder show better performance in most of the subject systems compared to the other clone detectors. On the other hand, Duplo, CLW-T1 shows the worst performance in most subject systems. Other tools show average performance considering individual ranking in different subject systems.  CLW-T1 and Duplo obtained the bottom position in the final rank list table. 

As our analysis was based on the clone grouping into class or pair provided by the clone detection tools, we found that clone detection tools' efficiency in suggesting cloned co-change candidates is mostly dependent on its effectiveness in making clone class/ pair. The tool which groups functionally similar clone fragments into a clone class/ pair effectively can perform well in successfully suggesting cloned co-change candidate(s). Different values of different clone detectors' accuracy indicate the difference in their efficiency in this research domain. 

\subsection{Answer to the \textbf{RQ2}}
\textbf{What are the deciding factors for the performance variance of different clone detectors in detecting cloned co-change candidates?} 

From the answer of our \textbf{RQ1}, we found a difference in performance for different clone detection tools in suggesting cloned co-change candidates. Although we selected all the promising clone detection tools reported in several studies, we found differences in their performance in detecting cloned co-change candidates. Such a scenario motivates us to find out the reason to answer this research question.

\begin{figure}
\centering
\includegraphics[width=\textwidth] {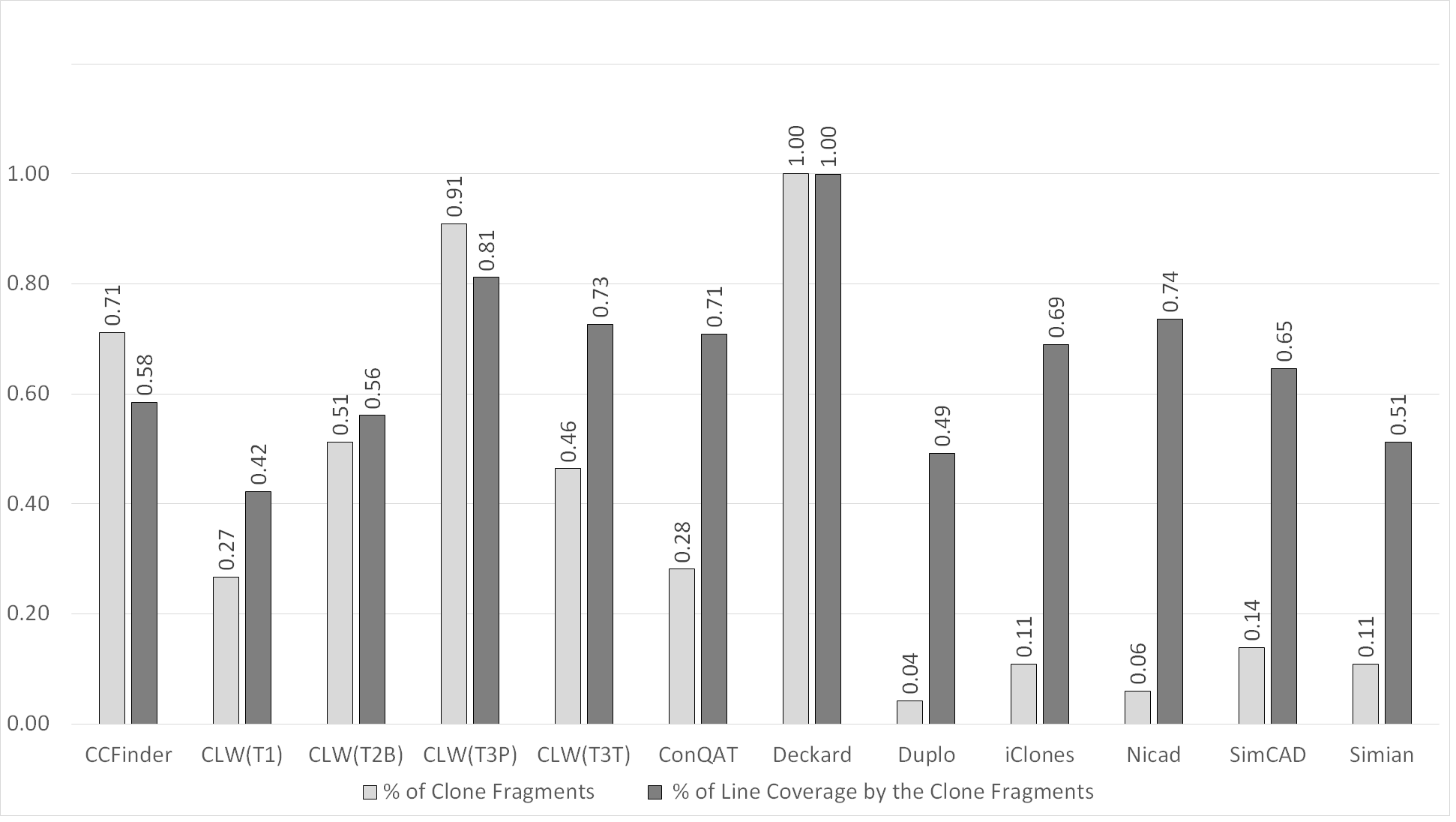}
\caption{Comparing unique line coverage by clone fragments and number of clone fragments from different clone detectors.}
\label{fig:AverageLineCoveredPerSS}
\end{figure}

We investigated the number of clone fragments and the number of unique lines covered by those clone fragments by all the 12 clone detectors from all the subject systems' revisions. Figure \ref{fig:AverageLineCoveredPerSS}  shows the comparison scenario of the number of clone fragments and lines covered by those clone fragments from different clone detectors. For better comparison, we bring the values on a single scale (between 0 and 1) where 0 and 1 represent the lowest and highest values, respectively, compared to all the clone detectors under comparison. Considering both, the number of clone fragments and the number of lines covered by those clone fragments from all the revisions of all the subject systems, if we order the clone detectors from the highest to the lowest, we find Deckard and CLW-T3P in the top of the list. CLW-T3T and CCFinder fall in the respective next position to provide the highest number of clone fragments and cover the highest number of unique lines in the source files. This scenario shows that a good clone detector can perform poorly in detecting cloned co-change candidates if it does not detect enough clone fragments and does not cover enough unique lines by those clone fragments in the source file. Though, earlier study \cite{Mondal-2014-PRC-2597073-2597104rankingCoChange} suggests that NiCad is an excellent clone detector in both of these cases, it falls at the bottom of the list. Even though NiCad performs very well in detecting clone fragments, it provides a lower number of clone fragments and also the lower number of line coverage by those clone fragments in the software systems. For that reason, while detecting the cloned co-change candidates, NiCad is showing lower F1~Score. The number of clone fragments and line coverage by those fragments seems to be an underlying factor behind the obtained comparison scenario of the clone detectors in predicting cloned co-change candidates. Some other factors such as (i) How clone fragments are overlapped with each other in a clone group? (ii) How is a clone detector determining the similarity among different fragments? (iii) What similarity score is used; it may also have an effect on the performance of predicting cloned co-change candidates. We plan to explore these factors in future studies.

\subsection{Answer to the \textbf{RQ3}}
\textbf{Do the source code processing techniques (Pattern/Token/Text-based processing) of the clone detection tools have any impact on their performance in detecting cloned co-change candidates?}

We can answer this research question by analyzing our final ranking of the clone detectors in Table \ref{tab:final-ranking-sum-of-ranks}. Top two clone detectors of the final rank list (Rank 1 and 2) work by extracting source code patterns from the codebase. CLW-T3P processes the source code terms by splitting into lines and then extracts code patterns. Deckard first generates vectors from the source file and then extracts a tree-like source pattern to match similarity among different source code fragments. The other five tools (Rank 3 to 7) in the rank list perform token-based source code processing, and the remaining five tools perform text-based source code processing for detecting clones from the source file. From this result, we can say that text-based clone detection tools are not suitable to be used in detecting cloned co-change candidates during software evolution. The tools which can detect more generalized clone fragments, especially pattern-based clone detectors, are perfect for detecting cloned co-change candidates. 

\subsection{Answer to the \textbf{RQ4}}
\textbf{Do clone detection tools designed for detecting different types of clones (Type 1, 2, 3) work differently in detecting cloned co-change candidates?}

From the final rank list of our clone detectors, we also find the relation of detected clone types with its ability to detect cloned co-change candidates. The rank list of clone detectors in Table \ref{tab:final-ranking-sum-of-ranks} shows that clone detecting tools such as CLW-T1, Duplo, which detects the only Type 1 clone, will not perform well in detecting cloned co-change candidates. On the other hand, tools such as CLW-T3P, CLW-T3T, Deckard, CCFinder perform very well in detecting cloned co-change candidates. The significance test results in Table \ref{tab:cochange-wilcoxon-rank-test} also show that four tools (two configurations of CloneWorks for Type-3, Deckard, and CCFinder) which perform significantly better than the other tools are also known as the clone detectors which detects Type-3 clones (Type-1 and Type-2 are also automatically included with Type-3 clones). Therefore, our findings suggest that we should choose those clone detectors to be used in detecting cloned co-change candidates that detect Type-3 clones with the other Type-1 and Type-2 clone fragments. 

\begin{figure}
\centering
\includegraphics[width=\textwidth] {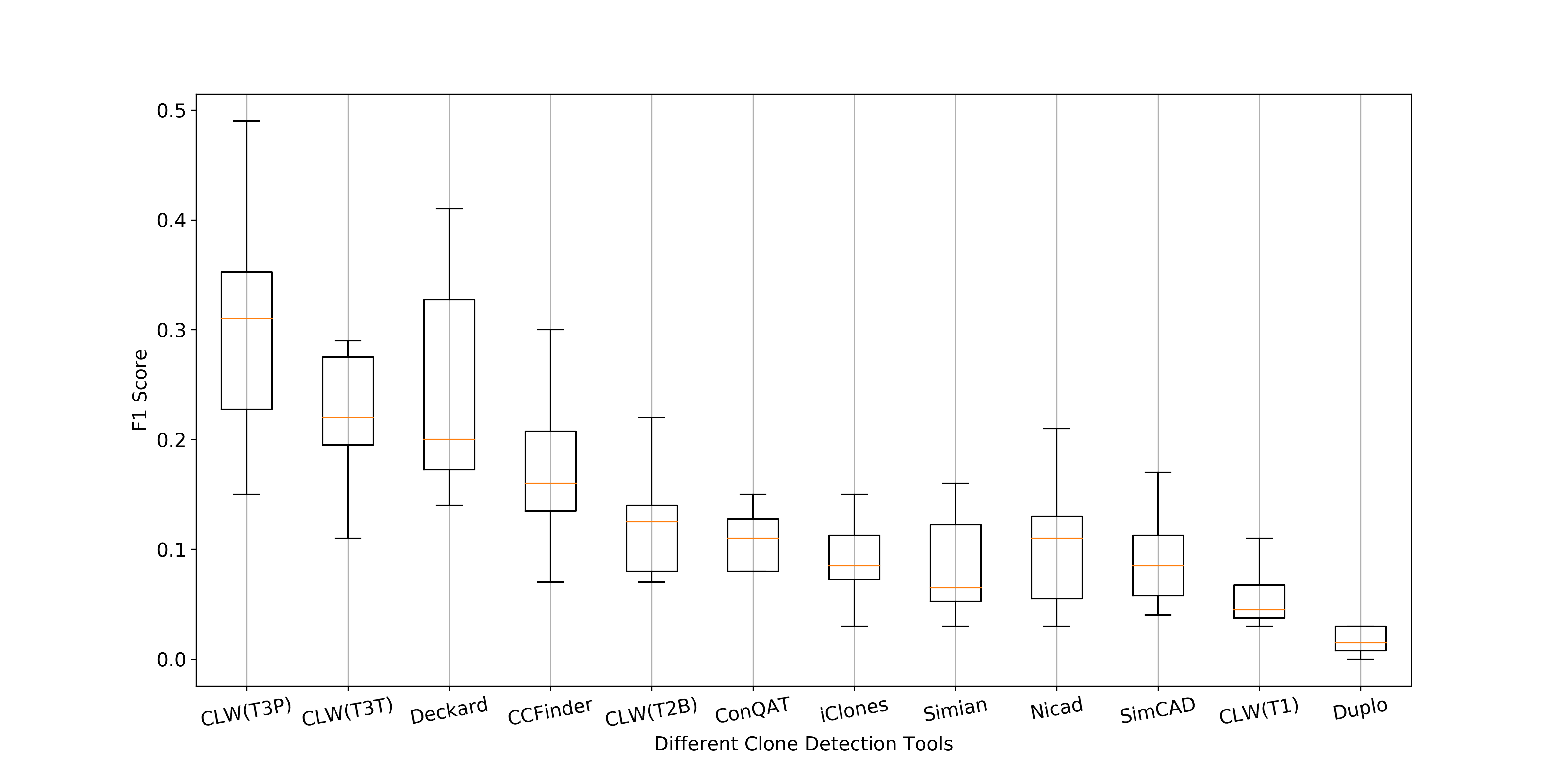}
\caption{Comparing Distribution of F1~Scores in Different Clone Detectors}
\label{fig:CochangeBoxFScoresRanked}
\end{figure}

\begin{table}[htbp]
\caption{\textsc{Wilcoxon Signed Rank Test (p<0.05)}}
\label{tab:cochange-wilcoxon-rank-test}
\centering
\begin{tabular}{|l|l|l|l|l|l|l|l|l|l|l|c|}
\hline
\multicolumn{1}{|c|}{\textbf{\begin{tabular}[c]{@{}c@{}}Tools in \\ Investigation\end{tabular}}} & \multicolumn{10}{c|}{\textbf{\begin{tabular}[c]{@{}c@{}}Significantly Better than \\ Tools (p\textless{}0.05)\end{tabular}}}                                    & \textbf{\begin{tabular}[c]{@{}c@{}}\# of \\ Tools\end{tabular}} \\ \hline \hline
\textbf{CLW-T3P}                                                                          & \multicolumn{10}{l|}{\begin{tabular}[c]{@{}l@{}}CLW-T3T, CCFinder, CLW-T2B, \\ ConQAT, iClones, Simian, Nicad, SimCAD,\\ CLW-T1, Duplo\end{tabular}} & 10                                                              \\ \hline
\textbf{CLW-T3T}                                                                            & \multicolumn{10}{l|}{\begin{tabular}[c]{@{}l@{}}CLW-T2B, ConQAT, iClones, Simian,\\ Nicad, SimCAD, CLW-T1, Duplo\end{tabular}}                            & 8                                                               \\ \hline
\textbf{Deckard}                                                                                 & \multicolumn{10}{l|}{\begin{tabular}[c]{@{}l@{}}CLW-T2B, iClones, Simian, Nicad, \\ SimCAD, CLW-T1, Duplo\end{tabular}}                                   & 7                                                               \\ \hline
\textbf{CCFinder}                                                                                & \multicolumn{10}{l|}{\begin{tabular}[c]{@{}l@{}}ConQAT, iClones, Simian, \\ SimCAD, CLW-T1, Duplo\end{tabular}}                                                & 6                                                               \\ \hline
\textbf{CLW-T2B}                                                                            & \multicolumn{10}{l|}{CLW-T1, Duplo}                                                                                                                            & 2                                                               \\ \hline
\textbf{ConQAT}                                                                                  & \multicolumn{10}{l|}{CLW-T1, Duplo}                                                                                                                            & 2                                                               \\ \hline
\textbf{iClones}                                                                                 & \multicolumn{10}{l|}{CLW-T1, Duplo}                                                                                                                            & 2                                                               \\ \hline
\textbf{Simian}                                                                                  & \multicolumn{10}{l|}{Duplo}                                                                                                                                     & 1                                                               \\ \hline
\textbf{Nicad}                                                                                   & \multicolumn{10}{l|}{Duplo}                                                                                                                                     & 1                                                               \\ \hline
\textbf{SimCAD}                                                                                  & \multicolumn{10}{l|}{CLW-T1, Duplo}                                                                                                                            & 2                                                               \\ \hline
\end{tabular}
\end{table}

\subsection{The Wilcoxon Signed-Rank Test:}
\label{sec-wilcoxon-singed-rank-test}
We performed The Wilcoxon Signed-Rank Test \cite{wilcoxon-signed-rank-test, wilcoxon-signed-rank-test-rosner} to verify the hypothesis that the F1~Scores of a tool which has obtained a higher rank in Table \ref{tab:final-ranking-sum-of-ranks} are significantly different (better) than the F1~Scores of the tools which have got lower ranks. Here, F1~Scores of each tool contains eight values obtained in all the eight subject systems. For instance, let us assume that we would like to examine whether the F1~Scores obtained by CLW-T3P are significantly better than the F1~Scores obtained by CLW-T3T. Thus, we take the sets of F1~Scores (see Table \ref{tab:detection-f1-score}) from both CLW-T3P and CLW-T3T, which will be then used to perform Wilcoxon Signed-Rank Test utilizing the SciPy library \cite{SciPy-NMeth2020} available in Python programming language. We did a significance test for each of the possible pairs from all the 12 clone detection tools in our investigation. 

A summary of the significant results at $p<0.05$ obtained from the significance test is given in Table \ref{tab:cochange-wilcoxon-rank-test}. The left-most column of this table contains the tool whose significance is to be tested, and the next column contains the name of the tools, each of them provides significantly different F1~Scores compared to the tool in the investigation. The right-most column of Table \ref{tab:cochange-wilcoxon-rank-test} shows the number of clone detectors whose F1~Scores are significantly different from the F1~Scores of the tool under investigation. Therefore, CLW-T3P provides significantly different F1~Scores than 10 other clone detectors (excluding Deckard). The distribution of F1~Scores in Figure \ref{fig:CochangeBoxFScoresRanked} also shows that majority of the F1~Score values of CLW-T3P lie above all the other clone detectors'  F1~Score values (except Deckard). Although some of the F1~Score values in CLW-T3P are above Deckard's values, those are not enough to make the result significantly different. This scenario clearly shows that CLW-T3P is significantly better than all the other clone detectors except Deckard. Similarly, from the following results of our significance test in Table \ref{tab:cochange-wilcoxon-rank-test}, we can see that F1~Scores of CLW-T3T are significantly better than the other eight clone detectors, F1~Scores of Deckard are significantly better than the other seven clone detectors, and F1~Scores of CCFinder are significantly better than the other six clone detectors. The following four tools (CLW-T2B, ConQAT, iClones, SimCAD) are significantly better than CLW-T1 and Duplo. Simian and NiCad are significantly better than only Duplo. The overall observation of the significance test result helps to conclude that for detecting cloned co-change candidates, CloneWorks Type-3 clone detection configuration can be an excellent choice, Deckard and CCFinder are also good choices. However, the other tools are not significantly better choices to detect cloned co-change candidates during software evolution. 

The distribution of F1~Scores in Figure \ref{fig:CochangeBoxFScoresRanked} also demonstrates the significance in performance differences of clone detectors used in this study. The clone detectors in this figure are sorted based on the final rank list shown in Table \ref{tab:final-ranking-sum-of-ranks}, where the ranks of the tools are presented from left to right (rank 1 to 12 in Table \ref{tab:final-ranking-sum-of-ranks}). This figure shows the clone detectors which got higher ranking in Table \ref{tab:final-ranking-sum-of-ranks} also have the higher values of F1~Scores compared to the tools which are below in the rank list. In this diagram, we can see that the F1~Scores of CloneWorks Type-3 Pattern have the most higher values, and Duplo has the most lower values in their respective distributions. Any two tools' performance will be significantly different from each other if they share a fewer common range of F1~Scores distribution. From the result of the significance test in Table \ref{tab:cochange-wilcoxon-rank-test}, we can see that Deckard is not significantly different from all the other three good clone detectors i.e. CLW-T3P, CLW-T3T, and CCFinder as they share most of the same range of values in the distribution. We can see a similar scenario for Simian and Nicad, e.g., though Simian and Nicad are above four and three other clone detectors, their F1~Scores are significantly better than only Duplo. Simian, Nicad, SimCad, CLW-T1 shares most of the same range of values in the distribution of F1~Scores. Therefore, they do not provide a significantly different result with each other. 

\section{Discussion}
\label{the-discussion}
There are two primary perspectives of managing code clones: (1) clone tracking and (2) clone refactoring. Our research principally focuses on the clone tracking perspective. A clone tracker's main task is to suggest similar cloned co-change candidates when a programmer attempts to change a code fragment. For suggesting cloned co-change candidates, a clone tracker depends on a clone detector. Our research compares 12 promising clone detectors based on their capabilities in suggesting cloned co-change candidates. 

According to our investigation, CloneWorks (Type-3 Pattern, and Type-3 Token), Deckard, and CCFinder are the most promising tools for suggesting such cloned co-change candidates based on the ranking we obtained in Table \ref{tab:final-ranking-sum-of-ranks} and the result of our significance test in Table \ref{tab:cochange-wilcoxon-rank-test}. Based on our overall observation, we can say that the performance of CloneWorks (Type-3 Pattern/ Token), Deckard, and CCFinder are much better compared to the other clone detection tools in detecting cloned co-change candidates during software evolution. As the clone classes/ pairs generated by different clone detectors played an essential role in our analysis, we can say that the clone detectors which can group similar clone fragments into a clone class/ pair efficiently will perform better in detecting cloned co-change candidates during the commit operation. From our findings, we can also say that the clone detectors which detect all the clone types such as Type-1, Type-2, and Type-3 clones can also perform well in detecting cloned co-change candidates. 

When a particular code fragment is changed, we apply the clone detectors to predict which other similar code fragments might need to be co-changed. However, some different fragments might also be changed together with the particular fragment. As we apply only clone detectors, we cannot consider those dissimilar co-change candidates in our research. We only apply our analysis to those change candidates whose co-changes are detected by at least one (out of 12) clone detection techniques in our investigation. We believe, removing the change candidates whose co-change is not detected by any of the clone detectors lead to a fair comparison among the clone detection tools.

\citet{Mondal:Association:Rules} show that combining Nicad as a clone detection tool with an association rule mining-based change impact analysis tool (Tarmaq \cite{TarmaqChangeImpact}) can significantly enhance the overall performance of detecting change impact set in software systems. As that study only included Nicad in their investigation, our study opens an opportunity to perform a few similar study other clone detectors which show better performance in our investigation. We find the few clone detectors such as CloneWorks, Deckard, CCFinder perform better in detecting cloned co-change candidates compared to Nicad. Therefore, including these tools in combination with Tarmaq might also serve better in change impact analysis. However, a practical evaluation is required to validate this assumption.  

\vspace{0.5cm}
In our research, we do not compare the clone detectors considering their clone detection efficiency. We instead compare the clone detection tools based on their ability to suggest cloned co-change candidates. Such a comparison of clone detectors focusing on a particular maintenance perspective was not made previously. However, suggesting both the cloned and dissimilar co-change candidates for a target program entity is a vital impact analysis \cite{book-change-impact} task during software evolution. Thus, through our research, we investigate which of the clone detectors can be helpful in change impact analysis to what extent. Findings from our study can identify which clone detector(s) can be promising for change impact analysis by finding the cloned co-change candidates. Furthermore, this study can also contribute to finding possible fixes of inconsistencies in software systems by analyzing historical discrepancies (due to missing the change in cloned co-change candidates) and their solutions (e.g., their fixing patterns).  

\section{Threats to Validity}
\label{the-threat-validity}
We use eight subject systems of diverse size and application domains written in C and Java programming languages and process thousands of revisions to rank 12 clone detection techniques for detecting cloned co-change candidates. The inclusion of more subject systems may increase the generalizability of the findings. The decision to select subject systems is based on the variety, popularity, used programming languages, and availability of a considerable number of revisions. We believe our results are not biased by our choice of subject systems and are distinguished from software maintenance perspectives. As the number of software revisions in our subject systems is extensive, our results should be generalizable to the other subject systems written in C or Java programming languages. Our results should also be generalizable to the programming languages whose structure is similar to C and Java, such as C++ or C\#. It might be possible that our results are not generalizable to other programming languages such as different scripting languages (e.g., Python, R, or PHP). We are detecting clone fragments using clone detectors and process that result to suggest cloned co-change candidates. As the clone detectors used in this study are established, and these are the most widely used tools for C and Java software projects \cite{sysReviewCloneDetectors}, their results might not be biased to any particular programming language. These tools are expected to work equally to all the subject systems written in C and Java programming languages. Therefore, we expect our results are generalizable to subject systems written in C and Java programming languages. However, in the future, we plan to investigate the generalizability of our technique to other programming languages by including a few more subject systems written in 
C\#, Go, Python, R, PHP, etc.

The parameters to detect clone fragments used in the 12 clone detectors may impact the comparison scenario in this study. However, we considered taking equivalent configurations (such as similarity threshold, length of tokens, renaming variables, splitting the source code lines, etc.) to minimize such impact. Therefore, we believe comparing clone detectors based on their detected cloned fragments and the cloned co-change candidate is made without influencing their clone detection configurations. 

We took both the fully and partially overlapped source code fragments as the intersecting code fragments from the results of clone detection tools and the cloned co-change candidates in our ground truth. Taking any other consideration to determine the overlap between two source code fragments might change the number of overall detected cloned co-change candidates. However, the change in the overlapping constraints should affect the results of all the clone detection tools and subject systems equally, and our overall comparison scenario of clone detection tools should still remain the same. We plan to practically verify different overlapping constraints in our future studies.  

Several code fragments might change together in a commit operation. While some of these fragments can be similar to one another, and some might be dissimilar. Similar code fragments co-change (e.g., change together) for ensuring consistency of the codebase. However, dissimilar code fragments can co-change because of their underlying dependencies, which could impact the generalization of this research outcome. As we aim to compare the clone detection tools, we wanted to discard the dissimilar co-change candidates from our consideration. If a co-change candidate was not detected as a true positive by any of the 12 clone detectors used in this study, we discarded the candidate. Excluding these changes, preparing the ground truth of cloned co-change candidates is essential for a fair comparison of the performance of clone detectors in this investigation. Using a different set of clone detectors might change the ground truth of cloned co-change candidates. Still, the ground truth of cloned co-change candidates in this investigation is accurate and complete for the set of 12 cloned detectors of this study. We believe that such a consideration is reasonable in our experiment aiming towards comparing clone detectors. Our findings may inspire more similar research with a different set of clone detectors in the future.

We did not consider a sliding window protocol that views some consecutive revisions in a single window to find cloned co-change candidates. There are some advantages of using the sliding window protocol. The sliding window protocol could provide a good result if some related cloned co-change candidates are scattered in two or more consecutive commits. As in this research, we are comparing 12 clone detectors. The effect of not including a sliding window protocol will be the same for each of the clone detectors, and thus, it does not have any impact on the final rank list of the clone detectors. However, we plan to apply this technique in our future research.

\section{Conclusion and Future Work}
\label{the-conclusion-cochange}
In this research, we compare different clone detection tools 
and investigate their performances in 
predicting
cloned co-change candidates during software evolution. 
We selected eight open-source subject systems written in C and Java for our analysis. Our final rank list in Table \ref{tab:final-ranking-sum-of-ranks} and summary of significance test results in Table \ref{tab:cochange-wilcoxon-rank-test} show that both the Pattern and Token configurations of the CloneWorks clone detection tool for detecting Type-3 clones perform significantly better compared to more than 72\% of the other clone detectors used in this study. Deckard and CCFinder are also better compared to more than 55\% of the other tools. CloneWorks (Type-2), ConQAT, iClones also show better performance than the other remaining tools. Our source codes, datasets, and other processed results are publicly available \cite{cochangeByClones} for researchers and practitioners to help continue and reproduce the results of this study. 

Although we determined some reasons for the better performance of Deckard, CloneWorks, and CCFinder in this extensive study, we plan to do some future work analyzing the internal mechanism of clone detection tools. We also want to find out how the change of these mechanisms affects the detection of cloned co-change candidates. In our future work, we also want to investigate the impact of different similarity scores of different clone detectors in finding cloned co-change candidates. We also want to include other software systems written in different programming languages (e.g.,  C\#, Go, Python, R, PHP). We want to consider a sliding window \cite{slidingWindow} protocol to consider more than one software revisions in a group to find cloned co-change candidates in the future extension of this study. 

\section*{Acknowledgment}
This research is supported by the Natural Sciences and Engineering Research Council of Canada (NSERC), and by a Canada First Research Excellence Fund (CFREF) grant coordinated by the Global Institute for Food Security (GIFS).

\bibliography{mybibfile}

\end{document}